\newif\ifbw
\bwfalse

\documentclass[aps,prl,preprint,showpacs,showkeys,superscriptaddress]{revtex4-1}
\usepackage{graphicx}
\usepackage{color}
\usepackage{tabularx}
\usepackage{bm}
\usepackage[english]{babel}

\begin{document}

\ifbw
  \newcommand{\FigureI}{fig1_BW.pdf}
  \newcommand{\FigureII}{fig2_BW.pdf}
  \newcommand{\FigureIII}{fig3_BW.pdf}
  \newcommand{\FigureIV}{fig4_BW.pdf}
\else
  \newcommand{\FigureI}{fig1.pdf}
  \newcommand{\FigureII}{fig2.pdf}
  \newcommand{\FigureIII}{fig3.pdf}
  \newcommand{\FigureIV}{fig4.pdf}
\fi

\title{All-in/all-out magnetic domains: X-ray diffraction imaging and magnetic field control}
\author{Samuel Tardif} 
\altaffiliation{current affiliation: CEA-Grenoble/INAC/SP2M/NRS, 17 rue des Martyrs, F-38054 Grenoble, France}
\email[]{samuel.tardif@gmail.com}
\affiliation{RIKEN, SPring-8 Center, Sayo, Hyogo 679-5148, Japan}
\author{Soshi Takeshita}
\affiliation{RIKEN, SPring-8 Center, Sayo, Hyogo 679-5148, Japan}
\author{Hiroyuki Ohsumi}
\affiliation{RIKEN, SPring-8 Center, Sayo, Hyogo 679-5148, Japan}
\author{Jun-ichi Yamaura}
\affiliation{MCES, Tokyo Institute of Technology, Kanagawa 226-8503, Japan}
\author{Daisuke Okuyama}
\affiliation{RIKEN, CEMS, Wako, Saitama 351-0198, Japan}
\author{Zenji Hiroi}
\affiliation{ISSP, University of Tokyo, Kashiwa 277-8581, Japan}
\author{Masaki Takata}
\affiliation{RIKEN, SPring-8 Center, Sayo, Hyogo 679-5148, Japan}
\affiliation{Department of Advanced Materials Science, University of Tokyo, Kashiwa 277-8561, Japan}
\affiliation{JASRI, SPring-8, Sayo, Hyogo 679-5148, Japan}
\author{Taka-hisa Arima}
\affiliation{RIKEN, SPring-8 Center, Sayo, Hyogo 679-5148, Japan}
\affiliation{RIKEN, CEMS, Wako, Saitama 351-0198, Japan}
\affiliation{Department of Advanced Materials Science, University of Tokyo, Kashiwa 277-8561, Japan}

\date{\today}


\begin{abstract}
Long-range non-collinear \textit{all-in/all-out} magnetic order has been directly observed for the first time in real space in the pyrochlore Cd$_2$Os$_2$O$_7$ using resonant magnetic microdiffraction at the Os L$_3$ edge. Two different antiferromagnetic domains related by time-reversal symmetry could be distinguished and have been mapped within the same single crystal. The two types of domains are akin to magnetic twins and were expected -- yet unobserved so far -- in the all-in/all-out model. Even though the magnetic domains are antiferromagnetic, we show that their distribution can be controlled using a magnetic field-cooling procedure. 
\end{abstract}

\maketitle

\begin{figure}
   \centering
   \includegraphics[width=88mm]{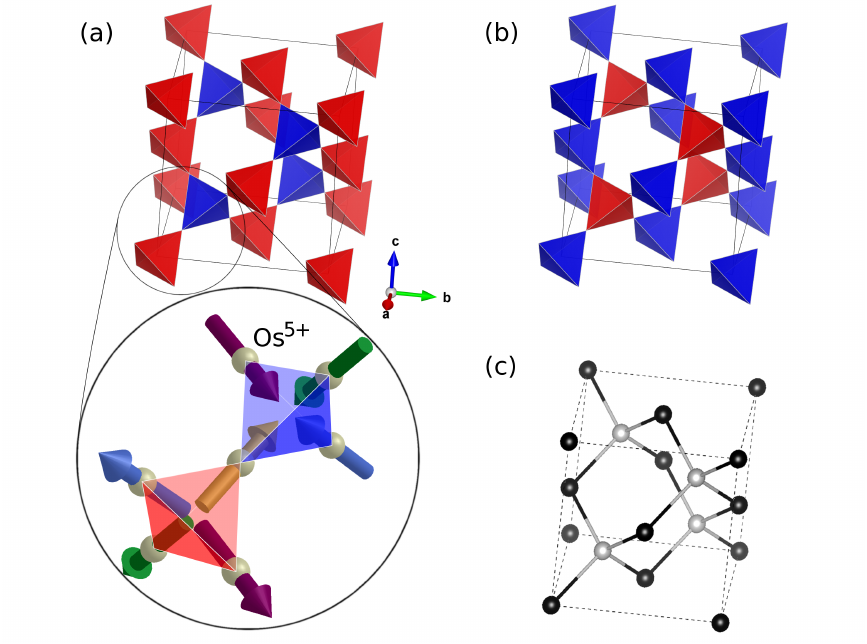} 
   \caption{
   (color online). (a) All-in/all-out and (b) all-out/all-in magnetic order on the pyrochlore lattice, both configurations are time-reversal symmetric of each other. 
   All Os spins are located at the vertices of the tetrahedra and point either towards the center of the blue (dark grey) tetrahedra or away from the center of the red (light grey) tetrahedra, as shown in the magnified region.
   (c) Equivalent zinc-blende lattice.} 
    \label{fig:Geo}
\end{figure}

The  pyrochlore lattice is a three-dimensional network of tetrahedra joined by their vertices and it has proven to be a rich host for new physics. For example, magnetic monopoles and Dirac strings have recently been reported in the so-called \textit{spin-ice} configuration \cite{morris_dirac_2009,fennell_magnetic_2009}, \textit{i.e.} when two of the four magnetic moments sitting at the vertices point toward the center of each tetrahedron while the other two moments point away from it (\textit{2-in/2-out}) \cite{bramwell_spin_2001,lacroix_introduction_2011}. 
When all the four magnetic moments point simultaneously either toward the center of the tetrahedron or away from it, \textit{i.e.} in the so-called \textit{all-in/all-out} (AIAO) order, the frustration is lifted and only two  magnetic configurations exist in the ground state: all-in/all-out and all-out/all-in. These two configurations are not equivalent and are related by time-reversal symmetry, as illustrated in Fig.\ref{fig:Geo}(a,b).  
The AIAO order is particularly interesting in many respects  in $5d$ transition metal oxides pyrochlores: the combination of the time-reversal symmetry breaking with strong spin-orbit coupling (due to the heavy magnetic element) and electron-electron correlations (evidenced by a metal-insulator transition) would be key ingredients to observe the predicted Weyl semimetal state \cite{wan_topological_2011}. 
Furthermore, the two time-reversal symmetric configurations can be seen as twins in a zinc-blende crystal of magnetic monopoles, or as differently oriented realisations of face-centered crystals of elementary magnetic skyrmions (Fig.\ref{fig:Geo}(c)) \cite{yu_real-space_2010}.
Finally, the AIAO order is equivalent to a Moessner \textit{pseudo-ferromagnet} \cite{moessner_relief_1998} and each configuration can be described in term of \textit{pseudo-orientation} (all-in/all-out and all-out/all-in being opposite pseudo-orientations). 
Domains with opposite pseudo-orientations are expected to display time-reversal odd magnetic properties, such as opposite parabolic magnetisation curves, or opposite linear magnetostriction, magnetocapacitance, and piezomagnetic effect \cite{arima_time-reversal_2013}.
Probing the magnetic state-dependent physical properties would therefore be highly desirable. However, in order to do so one would need (\textit{i}) to be able to match such measurements with those of the local magnetic order (\textit{i.e.} to be able to locally distinguish all-in/all-out from all-out/all-in) and (\textit{ii}) to be able to control the local magnetic order.

We propose here to use polarised resonant X-ray microdiffraction to distinguish between the two possible realisations of the AIAO order (\textit{all-in/all-out} and \textit{all-out/all-in}), \textit{i.e.} to measure experimentally the local pseudo-orientation. 
As a demonstration, we imaged simultaneously the two types of magnetic domains in Cd$_2$Os$_2$O$_7$ single crystals and report on a way to control their distribution.
The pyrochlore Cd$_2$Os$_2$O$_7$ is highly suited to investigate the properties of the AIAO order since the magnetic transition temperature is relatively high ($T_N=~225$~K). This temperature also correspond to a continuous metal-insulator transition (MIT) \cite{sleight_semiconductor-metal_1974}. 
The exact nature of the MIT and the closely related magnetic order of the low temperature insulating phase have recently been the focus of much interest \cite{mandrus_continuous_2001, reading_structure_2001,  singh_electronic_2002, padilla_searching_2002, harima_electronic_2002, koda_anomalous_2007, chern_spin_2011, matsuda_orbital_2011}.  
Using symmetry considerations and resonant X-ray diffraction, Yamaura \textit{et al.} recently showed that the spins on the 5\textit{d}$^{3}$ Os$^{5+}$ ions order according to the AIAO rule in the low temperature phase  \cite{yamaura_tetrahedral_2012} and this picture was further supported by various \textit{ab initio} calculations \cite{shinaoka_noncollinear_2012, bogdanov_magnetic_2013}.

Since X-ray diffraction can be site, element and electronic transition specific, it usually allows to determine the local absolute structural chirality \cite{bijvoet_determination_1951,tanaka_determination_2010,ohsumi_three-dimensional_2013}, local magnetic order \cite{lang_imaging_2004, kim_imaging_2005, evans_magnetic_2006} or local magnetic chirality \cite{hiraoka_spin-chiral_2011}, while circumventing the technical limitations of neutron scattering (non-local probe and limited by strong absorbers like Cd) or Lorentz transmission electron microscopy (requiring ultra-thin slabs).
In order to distinguish the all-in/all-out magnetic order from the all-out/all-in magnetic order, we made use of interference between the resonant magnetic and resonant non-magnetic terms of the structure factor.  
The resonant part of the structure factor of the space group-forbidden $0~0~4n+2$ reflections can be written expliciting its polarisation-dependence as a sum of two 2$\times$2 matrices between the conventional $(\mathbf{\sigma},\mathbf{\pi})$ and $(\mathbf{\sigma'},\mathbf{\pi'})$ bases:
\begin{eqnarray}
F^{(0~0~4n+2)} 
 &=& F_{ATS}\left|
   \begin{array}{cc} 
    \sin 2 \psi & \sin \theta \cos 2 \psi  \\ 
    -\sin \theta \cos 2 \psi  & \sin ^2 \theta \sin 2 \psi   
   \end{array} \right| \nonumber \\
 & &
   +
   F_{m}\left|
   \begin{array}{cc} 
   0 & \imath \sin \theta  \\
   \imath \sin \theta  & 0   
   \end{array} \right| ,  
\end{eqnarray}
where $F_{ATS}$ represent the non-magnetic term, related to the anisotropic tensor susceptibility (ATS) \cite{dmitrienko_forbidden_1983}, $F_{m}$ is the magnetic term, related to the magnetic order \cite{lovesey_x-ray_1996}, $\theta$ is the Bragg angle and $\psi$ is the azimuth angle, defined as the angle around the scattering vector such that $\psi = 0$ when $[100]$ is parallel to $\mathbf{k_{i}}+\mathbf{k_{f}}$, where $\mathbf{k_{i}}$ and $\mathbf{k_{f}}$ are the incident and diffracted wave vector, respectively.
As previously noted by Yamaura \textit{et al.} \cite{yamaura_tetrahedral_2012}, in the special experimental condition where $\psi = 45^\circ$, the scattering plane is parallel to the $(1\bar{1}0)$ plane (\textit{i.e.} the mirror plane broken by the AIAO order) and the off-diagonal components of the ATS vanish.
In this condition, the intensity $I^{\pm}$ scattered from an incoming beam with the elliptical polarisation $\mathbf{\varepsilon^{+}} = (\sigma_{0}^{+},\pi_{0}^{+})$ or $\mathbf{\varepsilon^{-}} = (\sigma_{0}^{-},\pi_{0}^{-})$ is respectively proportional to 
\begin{eqnarray} \label{eq:FR}
I^{\pm}  & \propto & \left\| \sigma_{0}^{\pm}F_{ATS} + \imath \pi_{0}^{\pm}  \sin \theta F_{m}\right\| ^2 \nonumber \\ 
         & &       + \left\| \imath \sigma_{0}^{\pm}\sin \theta F_{m} + \pi_{0}^{\pm}\sin ^2 \theta F_{ATS}\right\| ^2 .
\end{eqnarray}
We define the flipping ratio (FR) as the contrast in diffraction of an X-ray beam for two opposite handedness:
\begin{eqnarray}
\mathrm{FR} &=& \frac{I^{+}-I^{-}}{I^{+}+I^{-}}.
\end{eqnarray}
In the particular case of circular polarised light (right-handed and left-handed), the polarisation is conventionally described by  $(\sigma_{0}^{+},\pi_{0}^{+})=(1,-\imath)$, $(\sigma_{0}^{-},\pi_{0}^{-})=(1,\imath)$ respectively.
Accordingly the FR is explicitly given by
\begin{eqnarray}
\mathrm{FR} &=& \frac{2 r \cos\phi \sin\theta \left(1-\sin^{2}\theta\right)}{r^{2} \left(1+\sin^{4}\theta \right)+2 \sin^{2}\theta}
\end{eqnarray}
where $r = \left|F_{ATS}\right|/\left|F_{m}\right|$ and $\phi$ is the phase difference between $F_{ATS}$ and $F_{m}$.
Let us now look closely at the resonant terms: we note that the magnetic structure factor $F_{m}(\mathbf{Q})$ is proportional to $\mathbf{M}(\mathbf{Q})$, the $\mathbf{Q}$ Fourier component of the Fourier transform of the magnetisation $\mathbf{m}(\mathbf{r})$ (see Supplemental Material). At $\mathbf{Q} = 0~0~4n+2$, the sign of $\mathbf{M}(\mathbf{Q})$ is opposite in AIAO and AOAI domains. On the contrary, $F_{ATS}(\mathbf{Q})$ is only dependent on the local anisotropy which is the same for both AIAO and AOAI domains, therefore its sign is constant. As a result, $\phi$ takes the values of $0$ or $\pi$, \textit{i.e.} \textit{the sign of the FR is opposite in opposite types of domains} :  
\begin{eqnarray}
\mathrm{FR}_{AIAO} &=& -\mathrm{FR}_{AOAI}.
\end{eqnarray}
The FR can be calculated from successive measurements of the diffracted intensity of right-handed and left-handed circular polarised X-ray and used to distinguish the local magnetic pseudo-orientation.


We performed the FR measurements in high quality Cd$_2$Os$_2$O$_7$ single crystals, as described in Ref. \cite{yamaura_tetrahedral_2012}. 
The measurements were performed at the Os L$_{3}$ absorption edges (10.871 keV) and carried out on a magnetic diffraction setup (high resolution in reciprocal space) and on a microdiffraction setup (high resolution in real space), respectively in experimental hutch 1 and 4 of beamline BL19LXU at the SPring-8 synchrotron radiation source \cite{yabashi_design_2001,takeshita_preparation_????}. 

\begin{figure}
   \centering
   \includegraphics[width=88mm]{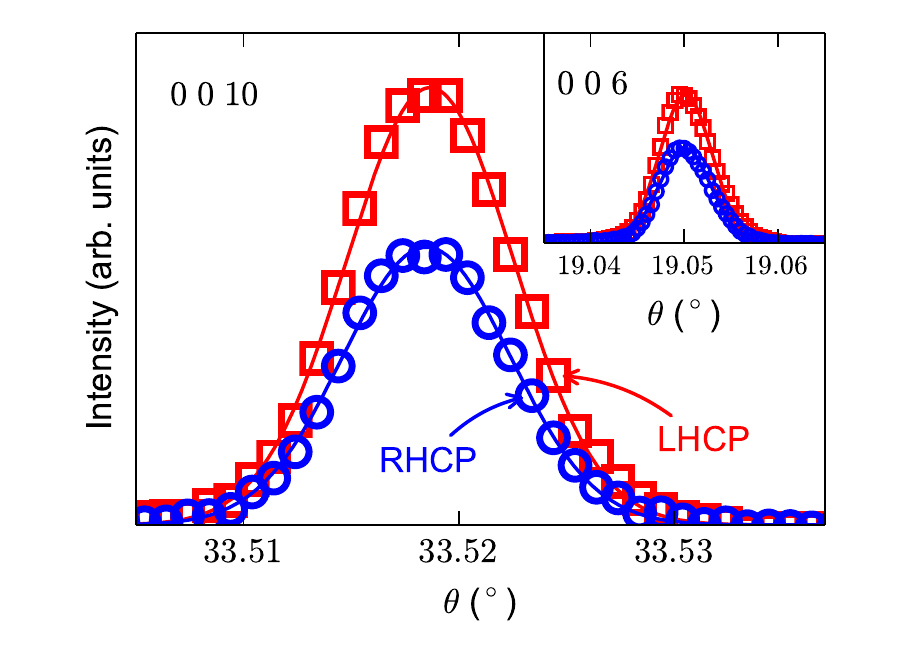}    
   \caption{
   (color online).
   Profiles of the forbidden $0~0~10$ reflection at the Os L$_3$ edge (10.871 keV) and at 10 K in the case of left-handed (LHCP) and right-handed (RHCP) circular polarised incident beam. 
   The inset shows the same measurement at the $0~0~6$ reflection. 
} 
    \label{fig:FR}
\end{figure}

After cooling the sample in a magnetic field along a $\langle111\rangle$ direction through the ordering temperature down to 10 K, the resonant magnetic X-ray diffraction was measured on the magnetic diffraction setup at space group-forbidden  $0~0~10$ and $0~0~6$ reflections for an incident X-ray beam with either left-handed circular polarisation or right-handed circular polarisation, as shown in Fig. \ref{fig:FR}. 
After correction from the incident intensity, a large difference was observed, amounting to a FR of about 22\%.  The sign of the FR was the same for both the $0~0~10$ and the $0~0~6$ reflections, which is consistent with the observation of either a single all-in/all-out (or all-out/all-in) domain or a strongly unbalanced population of all-in/all-out and all-out/all-in domains.

\begin{figure*}[h]
   \centering
   \includegraphics[width=170mm]{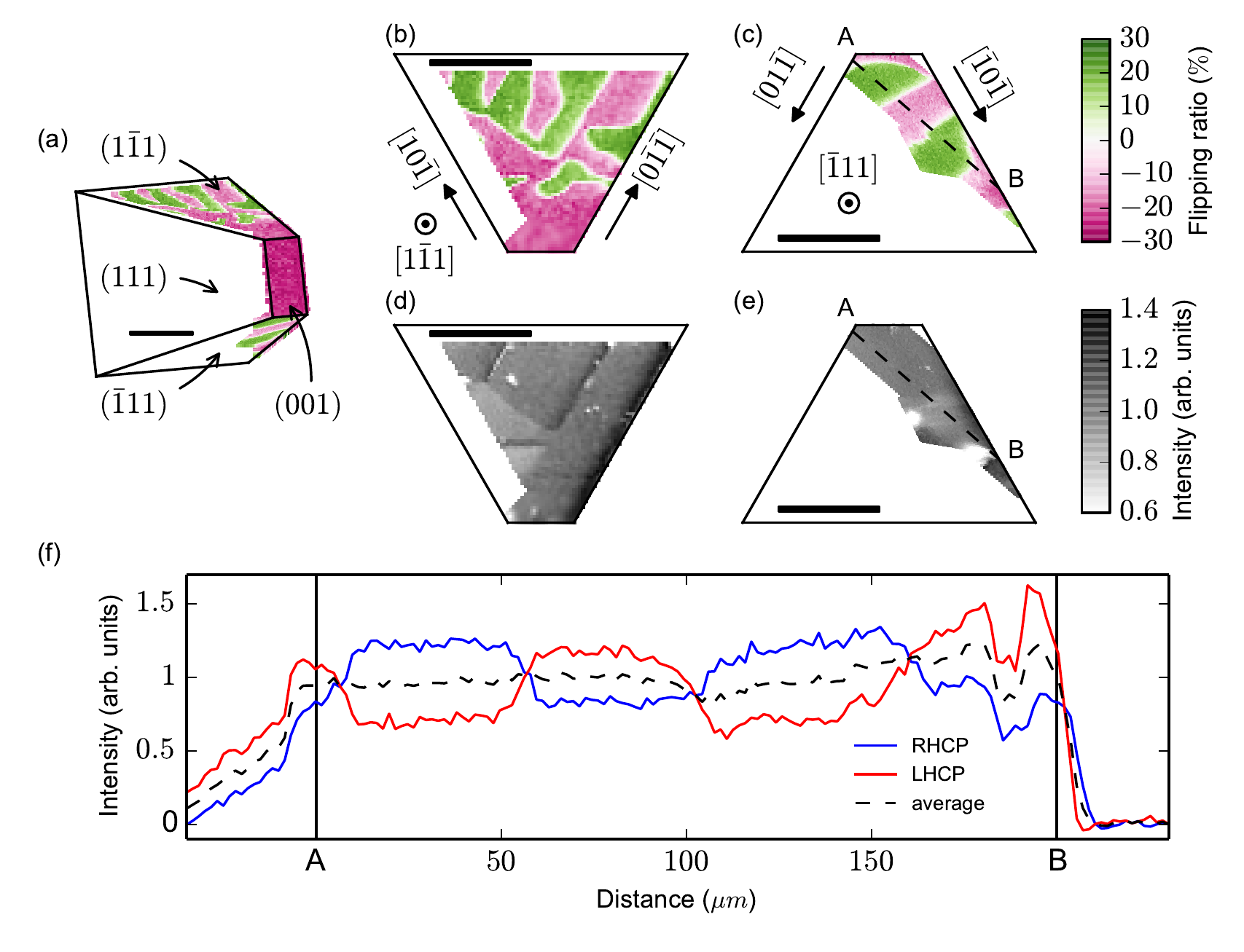} 
   \caption{
   (color online).
   (a) FR map at the $0~0~10$ reflection, as measured experimentally by rastering the sample in a plane perpendicular to the scattering vector. The ideal crystal is outlined and the orientation of the facets is shown. Opposite signs of the FR indicate opposite pseudo-orientations of the magnetic domains (AIAO \textit{vs} AOAI).
  (b,c) FR maps and (d,e) normalised intensity on the top $(1\bar{1}1)$ and bottom $(\bar{1}11)$ facets after correction of the projection, respectively. 
   The intensity is normalised to the average value on each facet to outline the deviation from the average structure. 
   All scale bars are $100~\mu m$ across. 
   (f) Line scan across the bottom $(\bar{1}11)$ facet, indicated by the AB dashed line in (c,e), for both right- and left-handed circular polarisation and their average. 
} 
    \label{fig:large_map}
\end{figure*}

\begin{figure}
   \centering
   \includegraphics[width=88mm]{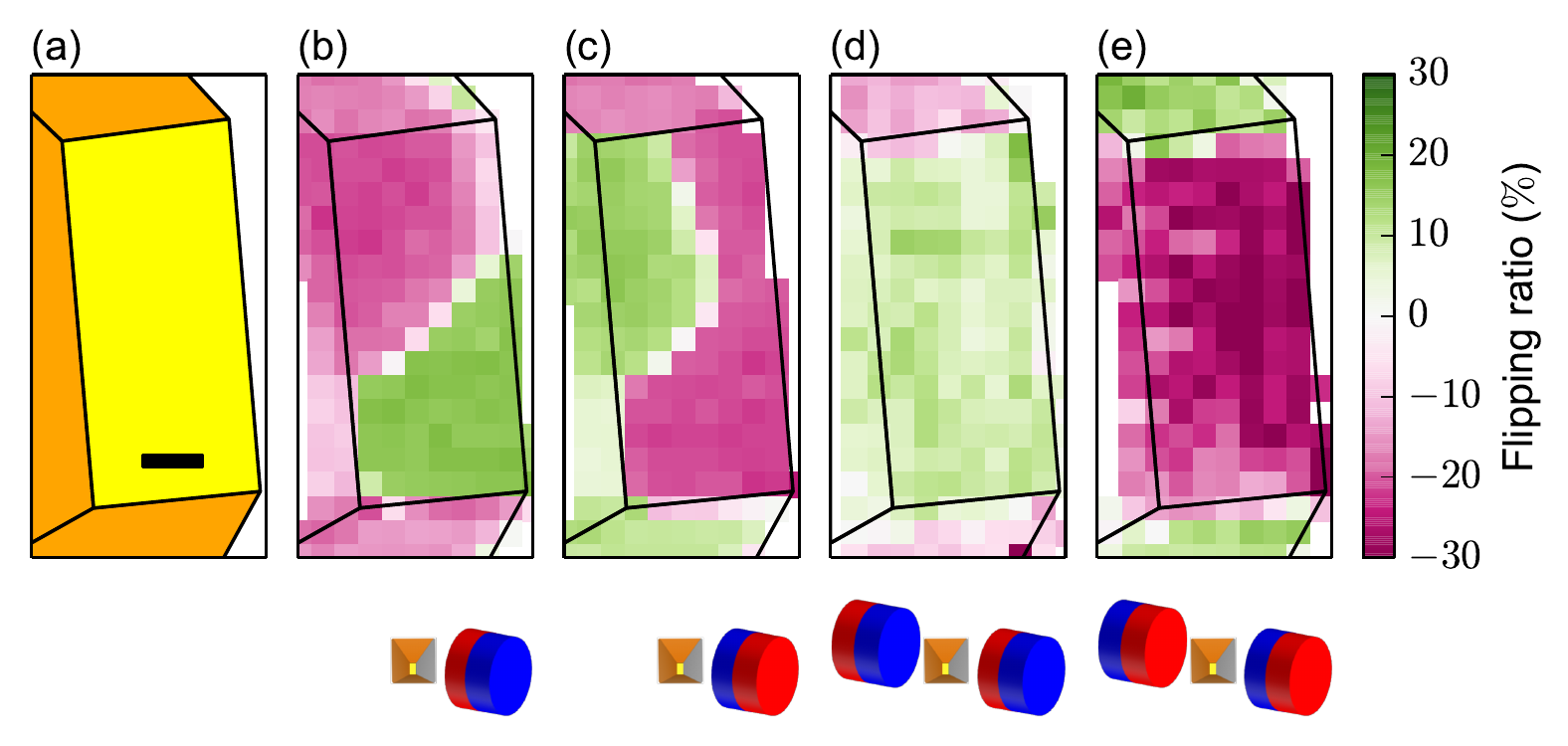} 
   \caption{
   (color online).
   (a) Sketch of the sample view in the geometry used, the central rectangle is the $(001)$ facet. 
   The scale bar is 25 $\mu m$ across.
   (b--e) Maps of the flipping ratio across the sample at the $0$~$0$~$10$ reflection at 100K for different field cooling conditions, as sketched in the lower panels.
    The North pole of the cylindrical permanent magnets is indicated in red (light grey), the magnetic field at the surface of the magnets is 0.4~T. 
    The surface of the magnet is also parallel to the $(\bar{1}\bar{1}1)$ facet, indicated in light grey in the sketch.}
    \label{fig:map_reversal}
\end{figure}

The local pseudo-orientation of the AIAO and AOAI domains was investigated with sub-micron resolution on the X-ray microdiffraction setup. 
The typical results of a mapping experiment with a probe size of 500$\times$500~nm$^2$ at the $0~0~10$ magnetic reflection and T$=100$~K are shown in Fig. \ref{fig:large_map}(a). The measurement goes over the center $(001)$ facet and the adjacent $\lbrace111\rbrace$ facets of the Cd$_2$Os$_2$O$_7$ crystal. 
In order to be able to compare the magnetic domain maps on different facets of the sample, we corrected the measured maps assuming a slightly misaligned ideal crystal, outlined in Fig. \ref{fig:large_map}(a). 
In Fig. \ref{fig:large_map}(a--c), one can see that in that configuration the center $(001)$ facet is a single domain with negative FR, while several magnetic domains and domain walls (DW) can be observed on the top and bottom $\lbrace111\rbrace$ facets along high symmetry orientations.
Both top and bottom $\lbrace111\rbrace$ facets show a stripe-like structure with a typical size in the range of a few tens of micrometers (Fig. \ref{fig:large_map}(b--c)).  
Some preferential orientations of the intersections of the DW with the facets are observed, \textit{e.g.} $[011]$, $[21\bar{1}]$, $[110]$, $[1\bar{1\bar{2}}]$, $[121]$ and $[10\bar{1}]$ on the top $(1\bar{1}1)$ facet and $[12\bar{1}]$, $[011]$, $[211]$ and $[1\bar{1}2]$ on the bottom $(\bar{1}11)$ facet.
Some orientations ($[011]$, $[21\bar{1}]$) are correlated with the distribution of the average intensity, as can be seen by comparing Figs. \ref{fig:large_map}(b) and \ref{fig:large_map}(d). 
This indicates that in those cases the local magnetic domain structure is probably pinned by crystal defects.
The other DW seem to occur at places where the crystal quality is homogeneous, as seen in Figs. \ref{fig:large_map}(c) and \ref{fig:large_map}(e), and further illustrated in Fig. \ref{fig:large_map}(f) where a linescan across the bottom facet shows the alternation of positive and negative FR domains while the average intensity remains constant.
We note that only two groups of equivalent DW planes could explain all the orientations of the intersections with the top and bottom facets simultaneously: the ``113'' group ($\{(113),(11\bar{3}),(1\bar{1}3),(\bar{1}13)\}$ and circular \textit{hkl} permutations) and the ``011'' group ($\{(011),(01\bar{1})\}$ and circular \textit{hkl} permutations). The frustation is lower in the 113 group than in the 011 group since there are respectively 1 and 4 frustated moment per unit cell at the DW position. As a result we would expect the 113 group to be more energetically favorable.  However, only the 011 group can account for the $[100]$ orientation observed in Fig. \ref{fig:map_reversal}(c), which makes it a more likely candidate for the DW orientation. This may indicate that some magnetic order arise within the frustrated magnetic domain interface.

In order to assess in more details the domain reversal observed previously, we performed the same mapping experiment after cooling the sample in a magnetic field. 
The magnetic domain distribution for different field-cooling configuration are displayed in Fig. \ref{fig:map_reversal}(b--e) while the outline of the corresponding ideal crystal is shown in Fig. \ref{fig:map_reversal}(a). 
In all cases the field was oriented along the $\pm[\bar{1}\bar{1}1]$ direction in the diffraction plane. 
One can see that for each pair of configuration, reversing the orientation of the magnetic field applied during cooling results in a reversal of the magnetic domain distribution. 
In the case of the single magnet (Fig. \ref{fig:map_reversal}(b--c)), there exist a DW across the center of the $(001)$ facet, oriented along the $[100]$ direction. 
Upon reversed magnetic field cooling, the domain distribution is reversed over most of the imaged area and the interface retains the same position while the upper right part of the image still remains with the same orientation. 
The possible origin of this asymmetry is the low reproducibility, homogeneity and/or magnetic field strength of the single magnet cooling procedure and the pinning of the DW by some crystalline defects.
On the opposite, using a two--magnet configuration yields a different domain distribution with single domains of opposite orientation over the $(001)$ facet and the top $(1\bar{1}1)$ and bottom $(\bar{1}11)$ facets respectively. 
Moreover, reversing the more homogeneous cooling field symmetrically reverses the domain distribution over all the imaged area (Fig. \ref{fig:map_reversal}(d--e)), demonstrating the reversability of the procedure.

The possible origins of the magnetic-field control of the AIAO and AOAI domains lie in the domain-dependent magnetic properties described by Arima \cite{arima_time-reversal_2013}. 
To the first order, any small strain that could be induced during the cooling independently of the applied magnetic field (for example due to the thermal expansion mismatch with the sample holder) will lowers the free energy of one of the two pseudo-orientations with respect to the other, in a given orientation of the applied magnetic field.
Reversing the magnetic field during cooling will in turn favor the opposite pseudo-orientation.
To the second order, the magnetisation possess a non-linear (parabolic) component, the sign of which is dependent on the pseudo-orientation, \textit{i.e.} opposite pseudo-orientations will be favored in opposite applied magnetic field.
Furthermore, we also suggest that $\lbrace111\rbrace$ surfaces may play a role. 
On such surfaces, the absence of the fourth Os ion in the tetrahedra results in uncompensated magnetic moments and one can expect a ferromagnetic order of the net magnetic moments on a perfect $\lbrace111\rbrace$ surface, perpendicular to the surface. 
It is possible that these uncompensated moments couple to the cooling field and set the orientation of the underlying all-in/all-out magnetic order during the cooldown. 
As a result, the ability to control the pseudo-orientation paves the way for the study of the magnetic domain-dependent properties, \textit{e.g.} in the case of $(111)$-oriented thin films.

In summary, we observed the long-range all-in/all-out magnetic order in the insulating phase of the pyrochlore Cd$_2$Os$_2$O$_7$ using polarised resonant magnetic X-ray diffraction. 
In particular, all-in/all-out magnetic domains could be distinguished from all-out/all-in domains according to the sign of the flipping ratio of left- and right-handed circularly polarised X-ray diffraction.
Both types of domains were shown to co-exist in single crystals, with sizes in the range of a few tens of microns.
Additionally, the pseudo-orientation of the domains could be easily controlled by a simple magnetic field cooling procedure, consistently with theoretical predictions in the AIAO model. 
The experimental techniques to image and control the local all-in/all-out order that we describe here are not limited to the particular Cd$_2$Os$_2$O$_7$ compound, but may be further be applied to any all-in/all-out-type pyrochlores. 


This work was funded by the Funding Program for World Leading Innovative R\&D on Science and Technology (FIRST) project ``Quantum Science on Strong Correlation'' (QS2C).
Experiments were performed under proposal 20130074 for the RIKEN beamlines at SPring-8.
We acknowledge H. Kodera and Y. Hattori for technical assistance during the preparation of the experiments as well as the staff of BL19LXU at SPring-8 for expert support. We also thank S. Tsujiuchi and N. Torimoto for their help during the experiments at BL19LXU.




%

\end{document}


\title{All-in/all-out magnetic domains: X-ray diffraction imaging and magnetic field control\\
SUPPLEMENTAL MATERIALS}
\author{Samuel Tardif} 
\altaffiliation{current affiliation: CEA-Grenoble/INAC/SP2M/NRS, 17 rue des Martyrs, F-38054 Grenoble, France}
\email[]{samuel.tardif@gmail.com}
\affiliation{RIKEN, SPring-8 Center, Sayo, Hyogo 679-5148, Japan}
\author{Soshi Takeshita}
\affiliation{RIKEN, SPring-8 Center, Sayo, Hyogo 679-5148, Japan}
\author{Hiroyuki Ohsumi}
\affiliation{RIKEN, SPring-8 Center, Sayo, Hyogo 679-5148, Japan}
\author{Junichi Yamaura}
\affiliation{MCES, Tokyo Institute of Technology, Kanagawa 226-8503, Japan}
\author{Daisuke Okuyama}
\affiliation{RIKEN, CEMS, Wako, Saitama 351-0198, Japan}
\author{Zenji Hiroi}
\affiliation{ISSP, University of Tokyo, Kashiwa 277-8581, Japan}
\author{Masaki Takata}
\affiliation{RIKEN, SPring-8 Center, Sayo, Hyogo 679-5148, Japan}
\affiliation{Department of Advanced Materials Science, University of Tokyo, Kashiwa 277-8561, Japan}
\affiliation{JASRI, SPring-8, Sayo, Hyogo 679-5148, Japan}
\author{Taka-hisa Arima}
\affiliation{RIKEN, SPring-8 Center, Sayo, Hyogo 679-5148, Japan}
\affiliation{RIKEN, CEMS, Wako, Saitama 351-0198, Japan}
\affiliation{Department of Advanced Materials Science, University of Tokyo, Kashiwa 277-8561, Japan}

\date{\today}

\maketitle


\section{Polarization dependence of the flipping ratio}
\begin{figure*}[b]
   \centering
   \includegraphics[width=88mm]{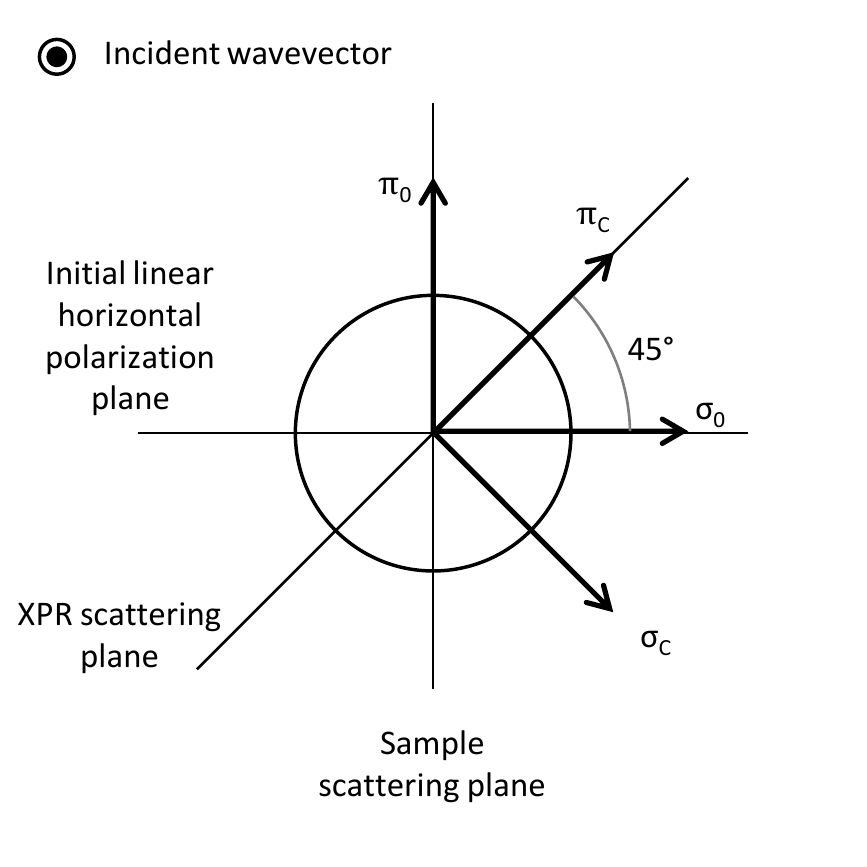} 
   \caption{
   Orientation of the polarization.
   Sketch of the orientation of the polarization for the magnetic diffraction setup. The initial incident beam from the undulator is linearly polarized in the horizontal plane. The crystal phase retarder introduces a phase difference between the $\sigma_c$ and $\pi_c$ components. The incident polarization on the sample, after the crystal phase retarder, is more easily described in the basis $(\sigma_0, \pi_0)$.
} 
    \label{fig:pol_geo}
\end{figure*}

The typical polarization of an X-ray beam produced by an undulator at a synchrotron facility is linear horizontal. In order to tune the incident polarization state, one can use a crystal in the Laue diffraction condition (\textit{i.e.} an X-ray crystal phase retarder, or XPR) and take advantage of dynamical diffraction effects.\cite{hirano_application_1997} Since the incident polarization is horizontal, the scattering plane of the XPR is rotated by 45 degree to allow equivalent diffraction in the $\sigma_c$ and $\pi_c$ channels, a necessary condition to achieve circular polarized light as illustrated in figure \ref{fig:pol_geo}.

\begin{figure*}
   \centering
   \includegraphics[width=88mm]{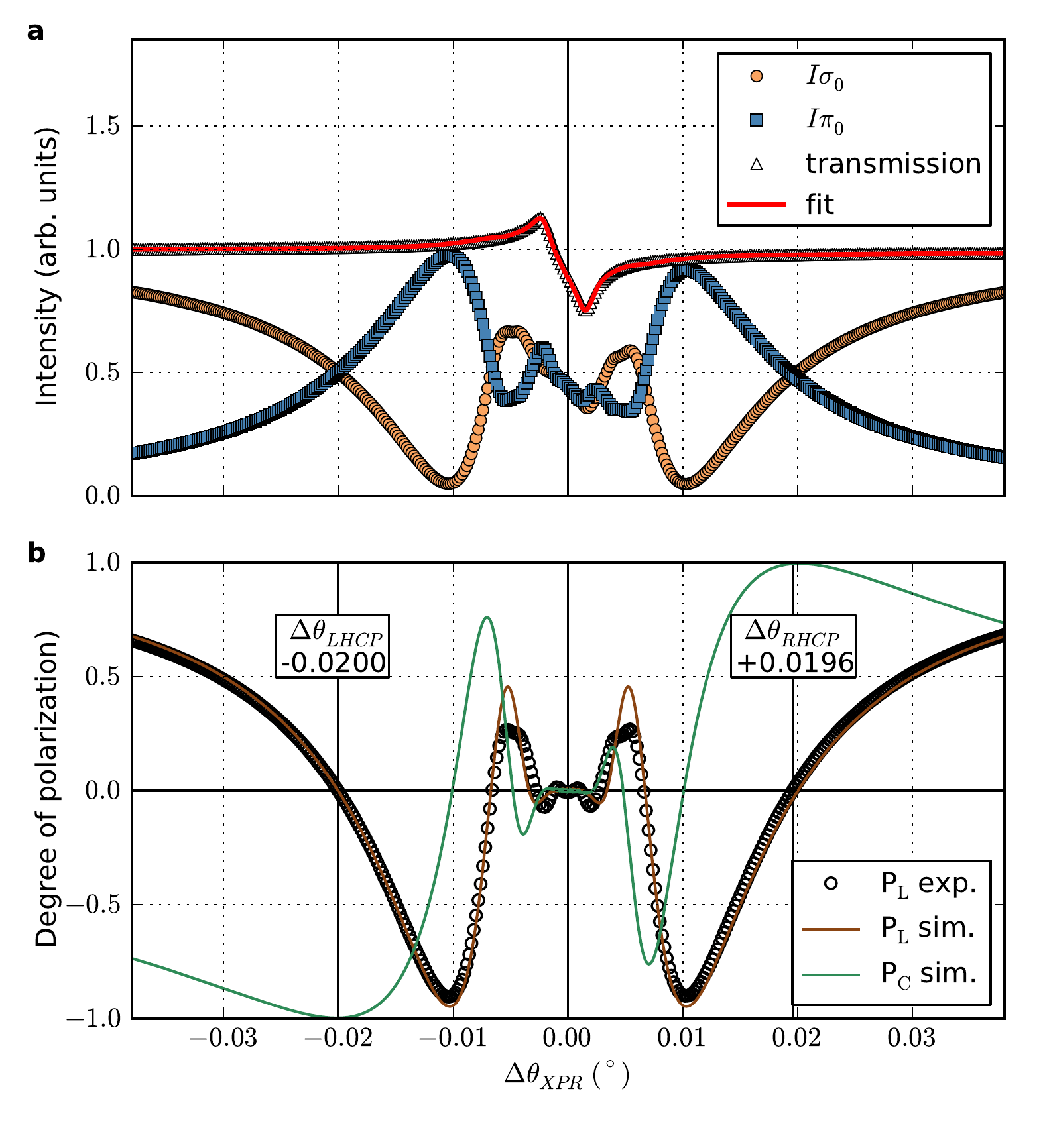} 
   \caption{
   Calibration of the X-ray polarization.
   (a), Experimental intensity of the $\sigma_{0}$ and $\pi_{0}$ components of the incident beam, as measured using a pyrolytic graphite analyzer crystal, as well as the total intensity transmitted through the XPR. The fit is a simple two parameters equation, where the parameters are the relative weight of each channel, \textit{i.e.} $0.997$ and $1.003$ for $\sigma_{0}$ and $\pi_{0}$, respectively. The small deviation from $1$ may be due to a small leaking of the cross polarization component during the polarization analysis, \textit{e.g.} due to the $2\theta$ angle of the analyzer crystal being different from $90^{\circ}$.
   (b), Experimentally measured degree of linear polarization $\mathrm{P_{L}}$, defined as $\mathrm{P_{L}} = \frac{I\sigma_{0}-I\pi_{0}}{I\sigma_{0}+I\pi_{0}}$, as well as the simulated $\mathrm{P_{L}}$ and corresponding degree of circular polarization $\mathrm{P_{C}}$.
} 
    \label{fig:pol_cal}
\end{figure*}

The XPR introduces a phase difference $\delta$ between the $\sigma_c$ and $\pi_c$ components of the downstream polarization of the beam (fig. \ref{fig:pol_geo}):
\begin{eqnarray} \label{eq:phase_shift}
\pi_c  &=& \sigma_c e^{-\imath \delta} ,
\end{eqnarray}
such that when $\delta=\pi/2$ or $\delta=-\pi/2$ the transmitted beam is right- or left-handed circularly polarized, respectively.
The phase difference is proportional to the inverse of the offset angle $\Delta\theta_{XPR}$ from the Bragg angle:
\begin{eqnarray} 
\delta &=& A /\Delta\theta_{XPR}  ,
\end{eqnarray}
where $A$ is a parameter depending on the Bragg angle of the reflection in the XPR , the corresponding crystal structure factor and the incident wavelength.\cite{hirano_application_1997} 
Since the offset for right-handed circular polarized light $\Delta\theta_{RHCP}$ can be experimentally measured, \textit{i.e.} when $\delta = \pi /2$ one can rewrite
\begin{eqnarray} \label{eq:phase_shift_angle}
\delta &=& \frac{\pi}{2} \frac{\Delta\theta_{RHCP}}{\Delta\theta_{XPR}} .
\end{eqnarray}
The polarization state downstream of the XPR is better described in the $\left(\sigma_0,\pi_0\right)$ basis, such that:
\begin{eqnarray}\label{eq:XPR_basis}
\pi_0  &=& \frac{\pi_c - \sigma_c}{\sqrt{2}} = \frac{e^{-\imath \delta}+1}{\sqrt{2}}\sigma_c  ,\\
\sigma_0  &=& \frac{\pi_c + \sigma_c}{\sqrt{2}} = \frac{e^{-\imath \delta}-1}{\sqrt{2}}\sigma_c  .
\end{eqnarray}
Substituting in equation \ref{M-eq:FR} in the Letter gives an analytical expression of the FR
\begin{eqnarray}\label{eq:fr_analytic}
\mathrm{FR}(\Delta\theta_{XPR}) = \pm \frac{ \cos \phi \sin( \frac{\pi}{2} \frac{\Delta\theta_{RHCP}}{\Delta\theta_{XPR}}) \sin \theta \left( 1 - \sin ^2 \theta \right) r}{r ^2 \left( \cos ^2 (\frac{\pi}{4} \frac{\Delta\theta_{RHCP}}{\Delta\theta_{XPR}}) + \sin ^4 \theta  \sin ^2 (\frac{\pi}{4} \frac{\Delta\theta_{RHCP}}{\Delta\theta_{XPR}}) \right) + \sin ^2 \theta } ,
\end{eqnarray}
where the only unknown parameters are the amplitude ratio $r=\left\| F_{m} \right\| / \left\| F_{ATS} \right\|$ and the phase difference $\phi$ between $F_m$ and $F_{ATS}$. 
Note that $\mathrm{FR}(\Delta\theta_{XPR})$ is obtained from a measurement of the diffraction with the XPR at $+\Delta\theta_{XPR}$ and $-\Delta\theta_{XPR}$ : due to the shape of the phase retarder crystal, the more precise absolute values of the offset angle for left-handed and right-handed circular polarization actually differ in the simulation of $\mathrm{P}_C$ by $1.4 \times 10^{-4}$ degree, which is lower than our experimental resolution ($2 \times 10^{-4}$ degree). 
It is worth mentioning however that small experimental misalignements can increase the difference between the absolute values of the negative and positive $\Delta\theta_{XPR}$. When mapping the flipping ratio over the sample, we used the calibrated values for circular polarization, such that the difference between $\left|+\Delta\theta_{XPR}\right|$ and $\left|-\Delta\theta_{XPR}\right|$ was between 0 and $4 \times 10^{-4}$ degree. A typical calibration curve is shown in figure \ref{fig:pol_cal}.

\section{Resonant magnetic diffraction from an all-in/all-out antiferromagnet}

The pyrochlore lattice of cadmium osmate belongs to the space group Fd\= 3m, where reflections $(hkl)$ are forbidden when $h,k,l$ are of mixed parity or when $h+k+l$ is not divisible by 4. In particular, $(0kl)$ reflections are forbidden when $k+l$ is not divisible by 4. Using the origin choice 2 in the International Tables of Crystallography, the osmium atoms are described by the 16\textit{c} Wyckoff sites $\mathbf{d}_{i} = (0,0,0)+\mathbf{t}, (\frac{3}{4},\frac{1}{4},\frac{1}{2})+\mathbf{t},(\frac{1}{4},\frac{1}{2},\frac{3}{4})+\mathbf{t},(\frac{1}{2},\frac{3}{4},\frac{1}{4})+\mathbf{t}$, where $\mathbf{t} \in \lbrace(0,0,0),(0,\frac{1}{2},\frac{1}{2}),(\frac{1}{2},0,\frac{1}{2}),(\frac{1}{2},\frac{1}{2},0)\rbrace$.

Since the all-in/all-out magnetic structure shares the same lattice as the underlying pyrochlore crystal ($\mathbf{q} = 0$ type), it also shares the same reciprocal lattice, \textit{i.e.} the magnetic diffraction can only occur at integer values of $(h,k,l)$. 
For the sake of simplicity, we neglect the corresponding Kronecker $\delta$ function in the following and restrict ourselves to the crystal unit cell. 

The resonant scattering term that we are interested in is the one that is linear with the local magnetic moment $\mathbf{m}$, the sign of which is opposite for the AIAO and AOAI domains. It is expressed as \cite{lovesey_x-ray_1996}
\begin{eqnarray}
\label{eq:fm}
f_{i,circ} &=& -\imath\left(\frac{3 \lambda}{8\pi ^{2}}\right)\left(\bm{\varepsilon '}\times \bm{\varepsilon }\right)\cdot \mathbf{m}_{i}\left[F_{-1}^{1}-F_{+1}^{1}\right]_{i}
\end{eqnarray} ,
where $\lambda$ is the wavelength of the incident x-ray beam, $\bm{\varepsilon '}$ and $\bm{\varepsilon}$ are the polarization vectors of the scattered and incident beam respectively, $\mathbf{m}_{i}$ is the unitary magnetic moment on the considered atom $i$ and $F_{M}^{L}$ are the resonant oscillator strength with $L=1$ and $M\in\left\lbrace-1,+1\right\rbrace$ in the dipole approximation.
As a result, the scattered amplitude is proportional to the scalar product of the cross-polarization term $\left(\bm{\varepsilon '}\times \bm{\varepsilon }\right)$ and the Fourier transform $\mathbf{M}(\mathbf{Q})$ of the distribution of the magnetic moments $\mathbf{m}$.

Let us first evaluate $\mathbf{M}(\mathbf{Q})$:
\begin{eqnarray}
\label{eq:FTM}
\mathbf{M}(\mathbf{Q}) 
 &=& \sum\limits_{ i}\mathrm{e}^{\imath2\pi\mathbf{Q}\cdot\mathbf{d}_{i}}\mathbf{m}_{i} ,
\end{eqnarray}
where $ \mathbf{d}_{i}$ is the position of atom $i$. We note that the magnetic moment direction distribution in a cubic all-in/all-out antiferromagnet can be described in the reduced basis $\left(\mathbf{\hat{a}}_{1},\mathbf{\hat{a}}_{2},\mathbf{\hat{a}}_{3} \right)$ of the unit cell as 
\begin{eqnarray}
\label{eq:maiao}
\mathbf{m}_{i,AIAO} 
  &=& \frac{\left| \mathbf{m} \right|}{\sqrt{3}} \sum\limits_{j=1,2,3} 
      \cos\left(4\pi\mathbf{r}_{i,j}\right)\mathbf{\hat{a}}_{j},
\end{eqnarray}
where $\mathbf{r}_{i,j} = \mathbf{d}_{i} \cdot \mathbf{\hat{a}}_{j}$.  Conversely, the magnetic moment direction distribution in the opposite domain (\textit{i.e.} all-out/all-in) is simply described by the opposite orientation of the magnetic moments:
\begin{equation}
\label{eq:maoai}
\mathbf{m}_{i,AOAI} = - \mathbf{m}_{i,AIAO} .
\end{equation}
We can now rewrite equation \ref{eq:FTM} for the all-in/all-out domain as
\begin{eqnarray}
\label{eq:FTM2}
\mathbf{M}(\mathbf{Q}) 
 &=& \frac{\left| \mathbf{m} \right|}{2\sqrt{3}} 
	 \sum\limits_{j=1,2,3}
	 \sum\limits_{ i} \left( \mathrm{e}^{ \imath 2\pi \left(\mathbf{Q}+2 \mathbf{r}_{i,j}\right)} + \mathrm{e}^{ -\imath 2\pi \left(\mathbf{Q}+2 \mathbf{r}_{i,j}\right)} \right) \mathbf{\hat{a}}_{j} .
\end{eqnarray}
Specifically, for a reflection $h,k,l=4n,4n,4n+2$, 
\begin{equation}
\label{eq:FTM3}
\mathbf{M}_{AIAO}(h=4n,k=4n,l=4n+2)) = \frac{16\left| \mathbf{m} \right|}{2\sqrt{3}}\mathbf{\hat{a}_{3}} .
\end{equation}
The sign of $\mathbf{M}_{AOAI}(\mathbf{Q})$ is the opposite for all-out/all-in domains. 

One can note that the amplitude of the Fourier transform of the magnetic moment orientation distribution is the same, even when the reflection is space-group forbidden. We can therefore choose the particular reflections $(0~0~4n+2)$. In this case $\mathbf{M}(\mathbf{Q})$ and the scattering vector $\mathbf{Q}$ are parallel to  $\mathbf{\hat{a}_{3}}$ (\textit{i.e.} the $\mathbf{c}$ axis) and it corresponds to the situation of specular reflection from the $\left( 0~0~1\right)$ facet. Such a situation is beneficial for the microscopy setup installed at BL19-LXU since the sample stage is moved parallel to the diffracting planes, \textit{i.e.} there is no geometric distortion in imaging the $\left( 0~0~1\right)$ facet. Furthermore, the facet edges provides reliable landmarks to compare pictures acquired at different times.
  
\begin{figure*}
   \centering
   \includegraphics[width=88mm]{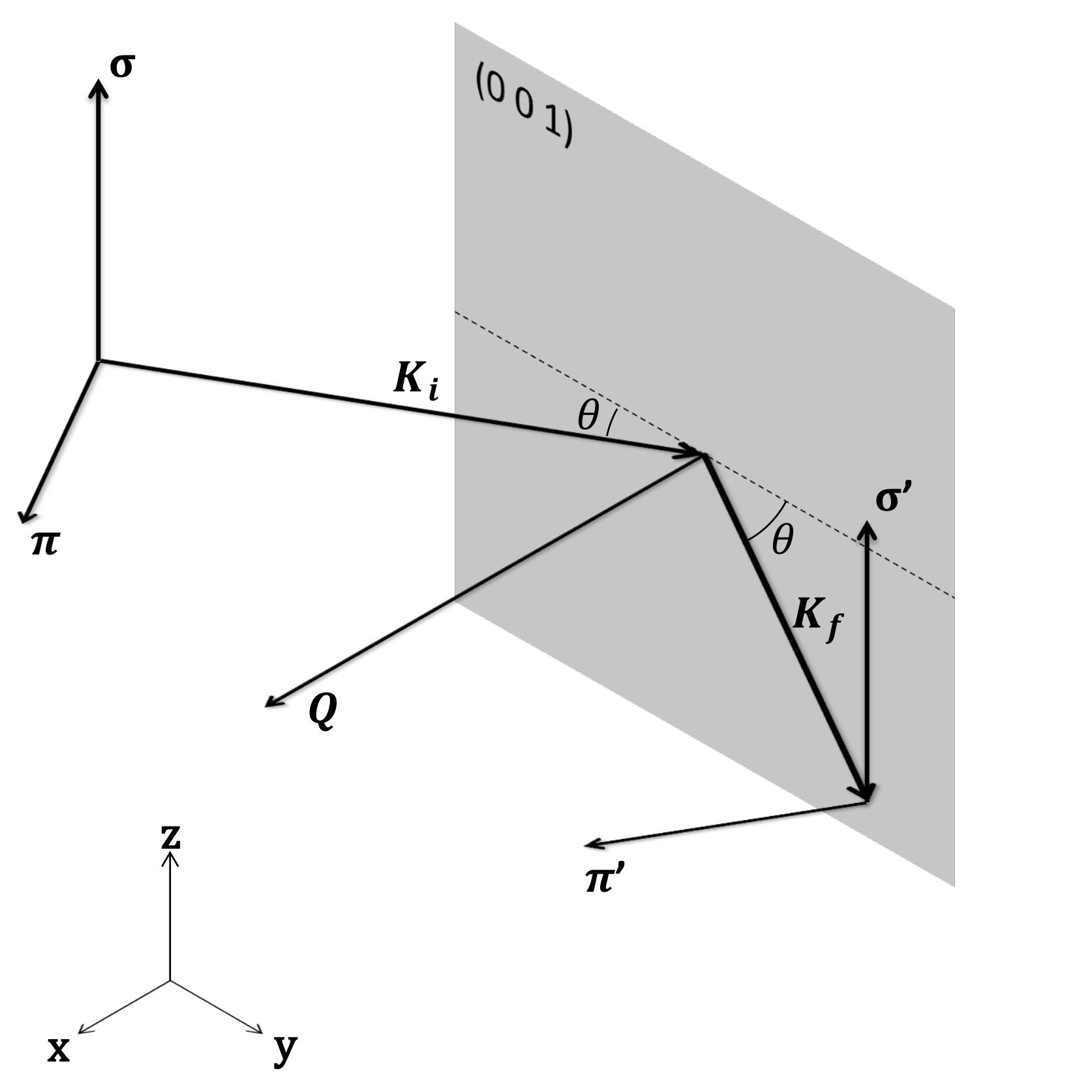} 
   \caption{
   Diffraction geometry at the microdiffraction setup.
   The $(0~0~1)$ facet lies in the $(\mathbf{y},\mathbf{z})$ plane. $\mathbf{K_i}$ and $\mathbf{K_f}$ are the incident and diffracted wave vectors, with incident and diffracted polarized states $\bm{\varepsilon } = \left( \bm{\sigma }, \bm{\pi } \right)$ and $\bm{\varepsilon '} = \left( \bm{\sigma '}, \bm{\pi '} \right)$ respectively. The scattering vector $\mathbf{Q}$ is along $\mathbf{x}$. Note that the scattering plane in this figure is rotated $90^{\circ}$ compared to figure \ref{fig:pol_geo}.
} 
    \label{fig:diff_geo}
\end{figure*}

We know turn to the cross-polarization term $\left(\bm{\varepsilon '}\times \bm{\varepsilon }\right)$. The diffraction geometry is sketched in figure \ref{fig:diff_geo}. The cross-polarization term can be written as a tensor linking the incoming polarization state $\bm{\varepsilon } = \left( \bm{\sigma }, \bm{\pi } \right)$ and the outgoing polarization state $\bm{\varepsilon '} = \left( \bm{\sigma '}, \bm{\pi '} \right)$. 
\begin{eqnarray}
\left(\bm{\varepsilon '}\times \bm{\varepsilon }\right) 
 &=& \label{eq:fcircMatrix} 
\bordermatrix{\text{ }          & \mathbf{\sigma}  &\mathbf{\pi} \cr
              \mathbf{\sigma '} & 0                &-\sin \theta\mathbf{x}+\cos \theta \mathbf{y} \cr
              \mathbf{\pi '}    & -\sin \theta \mathbf{x}-\cos \theta \mathbf{y} &  \sin 2\theta  \mathbf{z}}
\end{eqnarray} 
Since $\mathbf{M}(\mathbf{Q})$, $\mathbf{c}$ and $\mathbf{x}$ are parallel, considering equations \ref{eq:fm}, \ref{eq:FTM3} and \ref{eq:fcircMatrix}, the magnetic diffraction component for the all-in/all-out domain can be written as 
\begin{eqnarray}
\sum\limits_{ i}^{ }\mathrm{e}^{\imath2\pi\mathbf{Q}\cdot\mathbf{d}_{i}}f_{i,circ} 
 &=& 
 - \imath\left(\frac{3 \lambda}{8\pi ^{2}}\right) \mathbf{M}_{AIAO}(\mathbf{Q}) \cdot \mathbf{x} 
 \left(
 \begin{array}{cc} 
   0 & -\sin \theta \\
   -\sin \theta & 0    
 \end{array} 
 \right) \otimes
 \mathrm{FT}\left[F_{-1}^{1}-F_{+1}^{1}\right]\\
 &=& 
 F_m
 \left(
 \begin{array}{cc} 
   0 & \imath \sin \theta \\
   \imath \sin \theta & 0    
 \end{array} 
 \right) ,
\end{eqnarray} 
where the sign of the magnetic structure factor $F_m$ is opposite for AIAO and AOAI domain.


\section{Magnetic domain walls orientations}

\begin{figure*}
   \centering
   \includegraphics[width=170mm]{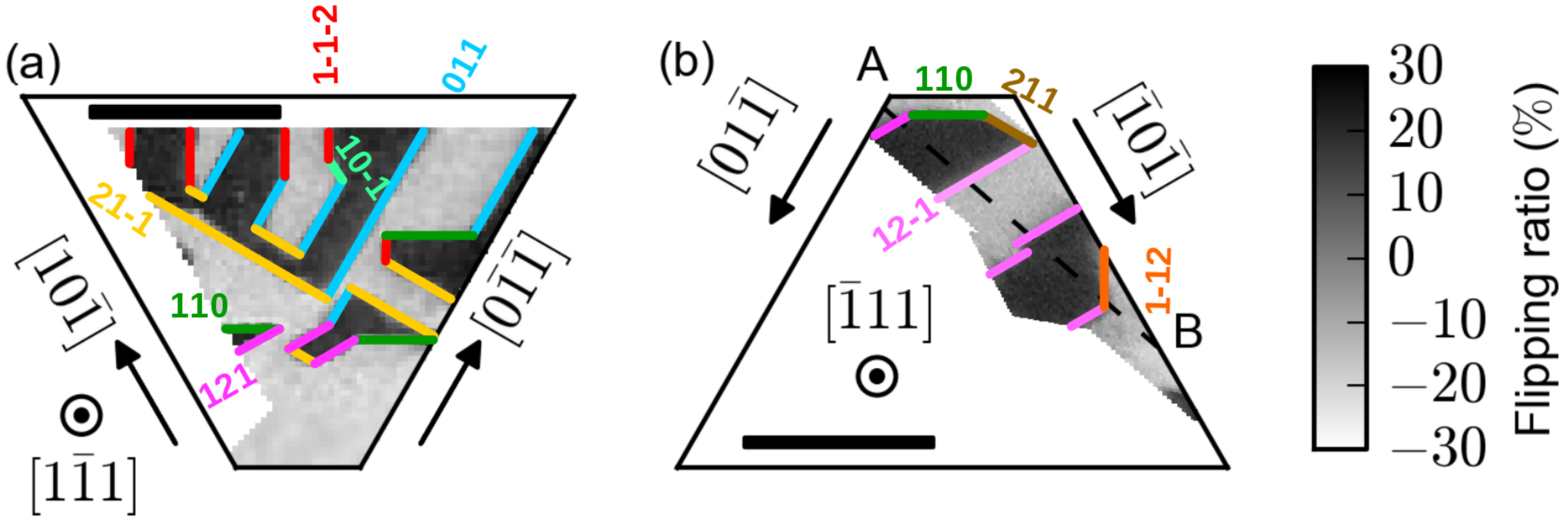} 
   \caption{
   FR maps on (a) the top $(1\bar{1}1)$ and (b) the bottom $(\bar{1}11)$ facets after correction of the projection and possible orientations of the intersections of the domain walls and the facets.
} 
    \label{fig:orientations}
\end{figure*}

In order to determine the orientation of the domain walls, we assigned the orientations of the intersections of the domain walls with the different facets to high symmetry directions, as shown in figure \ref{fig:orientations}.
As stated in the Letter, the only planes that can result in such intersections simultaneously on both facets are the ``113'' group ($\{(113),(11\bar{3}),(1\bar{1}3),(\bar{1}13)\}$ and circular \textit{hkl} permutations) and the ``011'' group ($\{(011),(01\bar{1})\}$ and circular \textit{hkl} permutations). 
The $(001)$ facet shows an additional two possible orientations of the domain wall intersection, as seen in figure \ref{M-fig:map_reversal}(b,c) in the Letter. The low datapoint count prevents a precise orientation determination on this facet. However one of the orientations may be $[\bar{1}10]$, which can come from either the 113 or the 011 group. The other orientation was more precisely measured (not shown here) to be $[100]$, which is only possible for the 011 group. 

\begin{figure*}
   \centering
   \includegraphics[width=170mm]{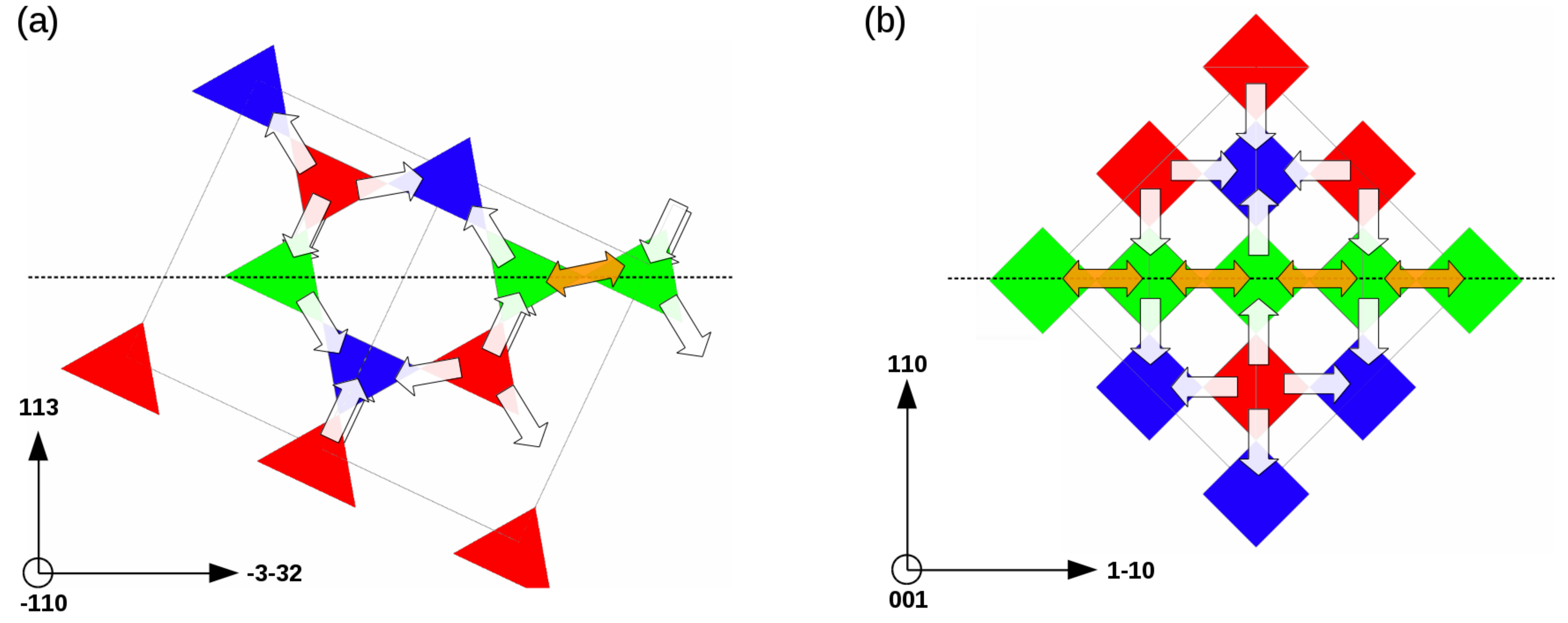} 
   \caption{
  (a) 113-type and (b) 011-type domain walls. Blue and red polygons are the all-in and all-out tetrahedra, respectively. White arrows indicate the magnetic moments in each all-in/all-out and all-out/all-in domains. Green polygons represent the tetrahedra at the domain interface, which are neither all-in nor all-out. Orange double arrows indicate the frustrated magnetic moments at the interface.
} 
    \label{fig:DW}
\end{figure*}

Both 113 and 011 types of domain walls are sketched in figure \ref{fig:DW}. The presence of the domain wall results in frustrated spins at the interface of the magnetic domains. There are respectively one and four frustrated spins per unit cell at the interface for 113-type and 011-type domain walls. One can therefore expect the properties of each type of interface to be quite different: in the 113 case the frustrated spins are independent while in the 011 case they are connected, which suggest the possibility of long-range ordering within the interface.


%